\documentclass[twocolumn]{aastex6}
\begin{document}


\title{WIYN Open Cluster Study. LXXII. A $uvbyCa$H$\beta$ CCD Analysis of the Metal-Deficient Open Cluster, NGC 2506}
\author{Barbara J. Anthony-Twarog}
\affil{Department of Physics and Astronomy, University of Kansas, Lawrence, KS 66045-7582, USA}
\email{bjat@ku.edu}
\author{Constantine P. Deliyannis}
\affil{Department of Astronomy, Indiana University, Bloomington, IN 47405-7105 }
\email{cdeliyan@indiana.edu}
\and
\author{Bruce A. Twarog}
\affil{Department of Physics and Astronomy, University of Kansas, Lawrence, KS 66045-7582, USA}
\email{btwarog@ku.edu}

\begin{abstract}
Precision $uvbyCa$H$\beta$ photometry of the metal-deficient, old open cluster, NGC 2506, is presented. The survey
covers an area $20\arcmin \times 20\arcmin$ and extends to $V\sim 18$ for $b-y$ and H$\beta$ and to $V \sim 17.0$ for
$c_1$ and $hk$. For $V$ brighter than 16.0, photometric scatter among the indices leads to the recovery of 6
known variables within the cluster core and 5 new variables in the outer $5\arcmin$ of the survey field. Proper motions, 
radial velocities, and precise multicolor indices are used to isolate a highly probable sample of cluster
members from the very rich color-magnitude diagram (CMD). From 257 highly probable members at the cluster turnoff, 
we derive a reddening estimate of $E(b-y) = 0.042 \pm 0.001$ ($E(B-V) = 0.058 \pm 0.001$), where the errors
refer to the internal standard errors of the mean. [Fe/H] is derived from the A/F dwarf 
members using both $m_1$ and $hk$, leading to [Fe/H]$ = -0.296 \pm 0.011$ (sem) and
$-0.317 \pm 0.004$ (sem), respectively.  The weighted average, heavily dominated by $hk$, is [Fe/H] $= -0.316 \pm 0.033$.
Based upon red giant members, we place an upper limit of $\pm 0.010$ on the variation in the reddening
across the face of the cluster. We also identify two dozen potential red giant cluster members outside the cluster
core. Victoria-Regina isochrones on the Str\"omgren system produce an 
excellent match to the cluster for an apparent modulus of $(m-M) = 12.75 \pm 0.1$  and an age of $1.85 \pm 0.05$ Gyr.

\end{abstract}


\section{Introduction}
NGC 2506 is an older open cluster which has long been grouped within the iconic class of metal-deficient clusters populating the
galactic anticenter just beyond the solar circle, as exemplified by NGC 2204, NGC 2243, NGC 2420, and Mel 66. A review of
the literature on these objects reveals an ongoing debate on just how old and how metal-deficient the individual members of the
class are, with clusters like NGC 2243 retaining consistently low metallicity ([Fe/H] $< -0.4$) as determined by photometry 
\citep{TW97, AT05}, medium-resolution spectroscopy \citep{FR02}, and high-resolution spectroscopy \citep{JA11, FR13}, 
while others like NGC 2420 generate [Fe/H] from $\sim-0.4$ to nearly solar, depending upon the technique adopted 
\citep{TW97, FR02, AT06, PA10, JP11}. The latter example is particularly relevant because, since the first 
comprehensive, complementary photometric \citep{MC81} and astrometric \citep{CA81} studies of NGC 2506, the
color-magnitude diagram (CMD) of NGC 2506 has exhibited features similar, though not identical, to those of NGC 2420.
Equally important, most recent moderate and high-dispersion spectroscopic work on a handful of giants has consistently 
shown that the cluster is more metal-poor than NGC 2420, with [Fe/H] $= -0.2$ or less \citep{CA04, MI11, RE12}, in agreement
with earlier analyses \citep{TW97, FR02}. As stellar isochrones and their calibration to standard photometric systems 
have improved over the last 30 years, the estimated age of NGC 2506 has declined steadily from greater 
than 3 Gyr, similar to M67, to less than 2 Gyr \citep{MC81, JP94, CA94, MA97, TW99, SA04, NE16}, making it comparable instead to 
NGC 3680, IC 4651, \citep{AT09} and NGC 752 \citep{TW15}.

The lower metallicity compared to NGC 752, NGC 3680, and especially IC 4651, in conjunction with the reduced age, makes NGC 2506 a 
key data point in delineating the roles of both parameters on the evolution of mixing/convection among stars 
from the main sequence through the giant branch, as well as the collateral impact on the atmospheric Li abundance. 
With astrometric membership \citep{CA81} extending at least three magnitudes below the level of a turnoff located
only $\sim$0.5 mag fainter than that of NGC 2420, NGC 2506 became an 
obvious choice for inclusion in an analysis of Li in stars from the tip of the giant branch to the level of 
the main sequence Li-dip in clusters covering a range in age and metallicity. Details on the scope of the project 
can be found in a number of papers discussing earlier results on the clusters NGC 3680 \citep{AT09}, 
NGC 6253 \citep{AT10, CU12}, NGC 752 \citep{MA13, TW15}, and NGC 6819 \citep{AT14, DE16} and will not be repeated. 
Suffice it to say that in any spectroscopic analysis, the validity of both star-to-star and cluster-to-cluster 
comparisons is ultimately tied to the consistency of the stellar parameters adopted in analyzing the spectra. 
This makes precise knowledge of the cluster reddening, stellar mass as defined by the cluster age and distance modulus, 
and an estimate of the cluster metallicity through photometry and spectroscopy critical to any evaluation of Li 
from the main sequence through the giant branch.

The purposes of this paper are to present estimates of the reddening and metallicity of NGC 2506 using precision CCD photometry
on the extended Str\"omgren ($uvbyCa$H$\beta$) system, to identify probable members from a region beyond the original 
astrometric survey of \citet{CA81}, and to derive an age and distance for the cluster through comparison with state-of-the-art 
isochrones on the Str\"omgren system. These data will serve as the basis for our spectroscopic analysis of the cluster from Hydra
spectroscopy extending from the giant branch to almost two magnitudes below the level of the turnoff.

The outline of the paper is as follows: Sec. 2 discusses the CCD observations and their reduction to the 
standard system for intermediate-band photometry; Sec. 3 uses the photometry, in conjunction with 
proper-motion membership, to identify and isolate probable cluster members which become the core data set 
for selecting single, main sequence stars for reddening, metallicity, age, and distance estimates in Sec. 4.  
Sec. 5 contains a summary of our conclusions.

\section{Observations and Data Reduction}

\subsection{Observations} 
NGC 2506 was observed using the 1.0 meter telescope operated by the SMARTS\footnote{http://www.astro.yale.edu/smarts} 
consortium at Cerro Tololo InterAmerican Observatory over the UT dates 16 - 22 December, 2011.  The telescope was 
equipped with  a STA $4000 \times 4000$ CCD camera\footnote{http://www.astronomy.ohio-state.edu/Y4KCam/detector} 
yielding a field of view 20\arcmin\ on a side at the Cassegrain focus of the 1.0m telescope. Filters owned by 
the University of Kansas were used to obtain images in each of the extended Str\"omgren system bandpasses 
($uvbyCa$ as well as the two H$\beta$ bandpasses). 
With the exception of 20/21 December, 2011, all of the nights were photometric.  
 
Images of NGC 2506 were obtained every night with a subset of the seven filters.  On three nights, frames to 
construct $V$, $b-y$, $hk$ and H$\beta$ were obtained; two nights were devoted to obtaining $uvby$ frames from 
which standard Str\"omgren indices could be constructed, with an additional two nights used to obtain $uvby$ 
and $Ca$ frames. In all, over 90 images of NGC 2506 were obtained in the course of the run, with total exposure 
times in minutes as follows:  $y$ (19) $b$ (25) $v$ (57)  $u$ (193) $\beta$ wide (26) $\beta$ narrow (97) $Ca$ (116). 
Field stars and clusters were observed at differing airmass to obtain extinction coefficients for all but the first 
night of the run.

Field star standards were observed on every photometric night, drawing from the catalogs of \citet{{OL83},{OL93},{OL94}} 
for Str\"omgren and H$\beta$ standards, using \citet{HA98} for additional H$\beta$ standard values. The catalog 
of \citet{TA95} was the principal source for $hk$ standards. Additional observations were obtained in several open 
clusters used as secondary standards, including M67 \citep{NI87}, NGC 2516 \citep{SN75} and NGC 2287 \citep{SC84}.   
In general, the secondary standard cluster fields would be used to delineate the slopes and color terms for the calibration 
equations, relying on the field star standards to set calibration equation zero-points for each photometric night.  

\floattable
\begin{deluxetable}{CCCCCh}
\tablecaption{Calibration coefficients \label{tab:table}}
\tablecolumns{6}
\tablenum{1}
\tablehead{
\colhead{Index} & \colhead{$\alpha$} & \colhead{$\beta$} & 
\colhead{$N_{ph}$\tablenotemark{a}} & \colhead{$N_{st}$\tablenotemark{b}} & \nocolhead{$\sigma$\tablenotemark{c}} 
}
\startdata
$V$ & 1.000 & 0.06 & 6 & 7-12 & 0.021 \\
$b-y(BD/RG)$ & 1.013 & & 6 & 5-10 & 0.009 \\
$b-y$ (RD) & 0.93 &  & 4 & 1-4 & 0.018 \\
$hk$ & 1.06 & & 4 & 7-8 & 0.019 \\
H$\beta$ & 1.09 & & 3 & 2-3 & 0.023 \\
$m_1$ RG & 0.923 & -0.046 & 3 & 2-4 & 0.071 \\
$c_1$ RG & 0.854 & 0.056 & 3 & 2-4 & 0.063 \\
$m_1$ BD & 1.010 & 0.04 & 3 & 3-7 & 0.082 \\
$c_1$ BD & 1.060 & 0.12 & 3 & 3-7 & 0.070 \\
\enddata
\tablenotetext{a}{Number of photometric nights on which standards were observed for this index.}
\tablenotetext{b}{Range of number of field star standards observed for this index each photometric night.}
\tablecomments{For each index $X_i$, the calibrated value is $\alpha X_i + \beta (b-y)_{instr} + \gamma$ }
\end{deluxetable}

Bias frames were obtained before every night's observations. Flat field correction was accomplished using a master set 
of sky flats assembled over the course of the entire run. Basic processing steps (bias subtraction and flat-field correction) 
were completed in IRAF, using the {\it y4k} scripts developed and described by Phil 
Massey\footnote{http://www2.lowell.edu/users/massey/obins/y4kcamred.html}.  DAOPHOT routines within 
the IRAF suite were used to obtain {\it psf}-based indices for every star on every NGC 2506 frame; our procedures 
for measuring and merging high-precision {\it psf}-based magnitudes and indices are fully described in \citet{AT00}.  

If internal standards are not available, final calibration of this precise set of magnitudes and indices must 
be anchored to airmass-corrected flux-based magnitudes of field star and cluster standards. To accommodate seeing 
variations, every aperture measurement used for calibration purposes is determined within an aperture radius set 
to five times the full width half maximum of the stars on that frame.   Similar measurements were obtained for 
a subset of the less crowded stars in NGC 2506 for each night, to facilitate the determination of aperture corrections.  

Using common slopes for each index, separate calibration equation zero-points may be determined for each relevant 
index for each photometric night. Slope and color terms used for the calibration equations are summarized in Table 1. 
Figure 1 illustrates the internal precision obtained in each index as a function of magnitude.

\begin{figure}
\figurenum{1}
\plotone{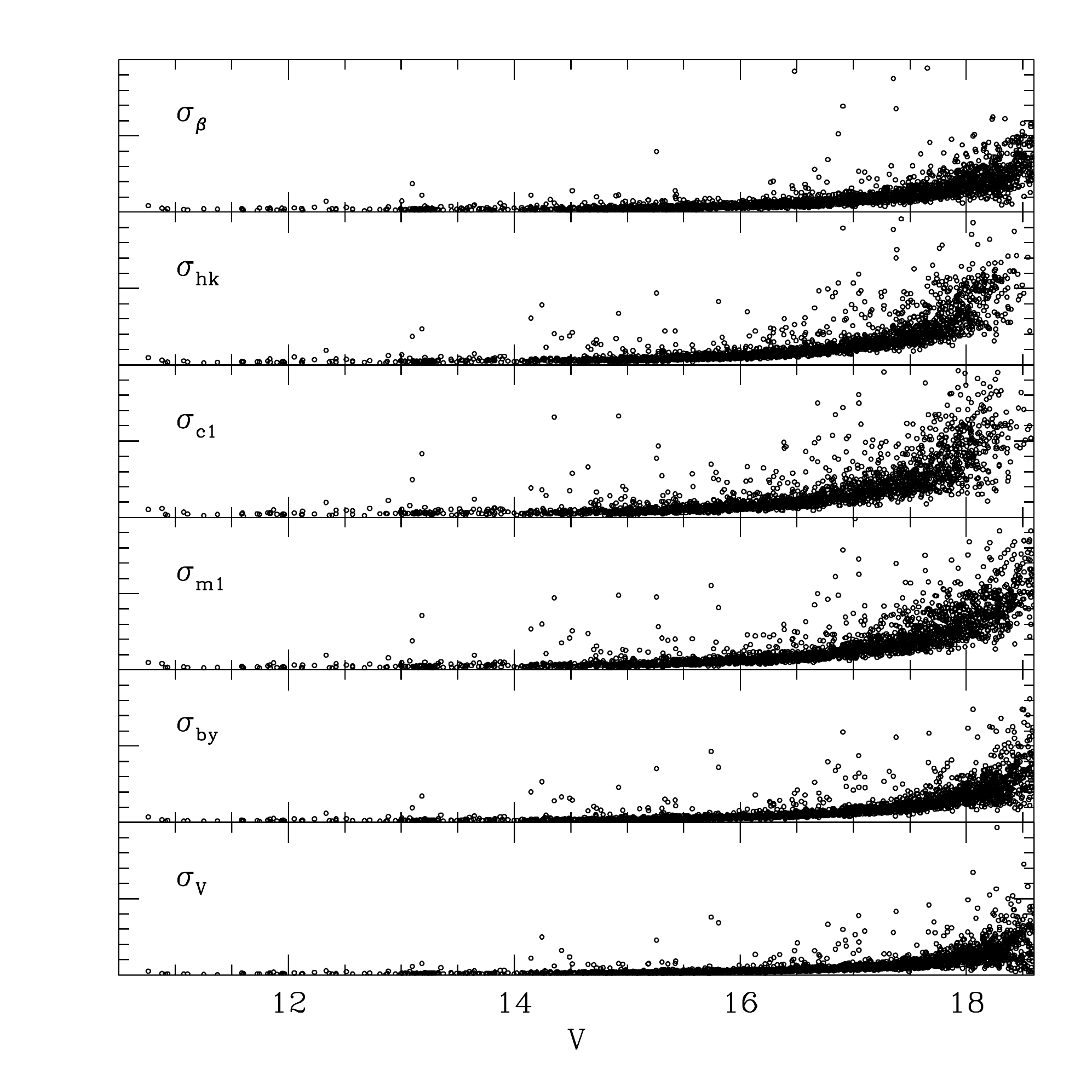}
\caption{Standard errors of the mean magnitude or index for stars in NGC 2506, as a function of $V$ magnitude. Major tick marks
on the vertical scale are in units of 0.010 mag in all plots. \label{fig:fig1}}
\end{figure}

Table 1 includes distinct calibration equations for some indices and stellar classes.  It was possible to 
determine a separate calibration equation for the $b-y$ colors of cooler dwarfs, distinguishable from the 
calibration equation used for dwarfs bluer than $(b-y)_0 \sim 0.42$ and red giants, even though our photometry 
in NGC 2506 doesn't really go deep enough for stars in this calibration class to be likely cluster members. More critical is the separation of calibration
equations for $m_1$ and $c_1$ for bluer, unevolved stars as opposed to cool giants and subgiants.  Aside from the 
challenge of determining these separate calibration equation segments, it is necessary to have some means of 
determining from uncalibrated photometry, to which class a star belongs.  Color alone adequately distinguishes 
blue dwarfs from red stars; separating redder evolved and unevolved stars is possible in the $hk, (b-y)$ plane as 
illustrated in \citet{TW15} for the case of NGC 752. 

Table 2 describes calibrated photometry for 2229 stars; the typeset version of the table includes enough lines to show the form and
content of the larger table. Stars with index values set to 9.999 fail to meet a criterion requiring at least
two observations in every relevant filter. Because of
the rapid increase in errors at fainter magnitudes, the Table has been restricted to stars with $V$ $\leqq$ 18.0. 
The Table includes X,Y pixel coordinates on the WEBDA system, RA(2000) and DEC(2000), followed by WEBDA ID and membership probability from \citet{CA81}, if available.
The photometry sequence is $V$, $b-y$, $m_1$, $c_1$, $hk$, and H$\beta$, followed by the calibration classification
B for blue dwarf, G for cool giant and D for cool dwarf. Finally, the standard errors of the mean are presented for each index using the
same sequence as the photometry, followed by the number of measures of the star in each filter.

\floattable
\begin{deluxetable}{rrrrccrrrrrrrrrrrrrrrrrrrrrrrr}
\tablecaption{Str\"omgren Photometry in NGC 2506}
\tablecolumns{26}
\tablenum{2}
\tabletypesize\tiny
\setlength{\tabcolsep} {0.03in}
\tablehead{
\colhead{$X_{web}$} & \colhead{$Y_{web}$} & \colhead{$\alpha(2000)$} & \colhead{$\delta(2000)$} & \colhead{WEBDA} & \colhead{Membership} & 
\colhead{$V$} & \colhead{$b-y$} & \colhead{$m_1$} & \colhead{$c_1$} & \colhead{$hk$} & 
\colhead{H$\beta$} & \colhead{$\sigma_V$} & \colhead{$\sigma_{by}$} & \colhead{$\sigma_{m1}$} & \colhead{$\sigma_{c1}$} & \colhead{$\sigma_{hk}$} & \colhead{$\sigma_{\beta}$} & 
\colhead{Class} & \colhead{$N_y$} & \colhead{$N_b$} & \colhead{$N_v$} & \colhead{$N_u$} & \colhead{$N_{Ca}$} & \colhead{$N_n$} & \colhead{$N_w$}
}
\startdata 
  434.07& -98.74& 120.124924& -10.799914&    &     & 10.761&  1.221&  0.544&  0.162&  1.581 & 2.686 & 0.005& 0.007& 0.010& 0.010& 0.009& 0.009&G &   20& 14 & 9& 14& 14 & 8 & 7 \\
   48.67& 522.25& 120.021011& -10.630401&    &     & 10.883&  1.041&  0.806& -0.139&  1.712 & 2.573 & 0.002& 0.004& 0.008& 0.012& 0.006& 0.005&G &   22& 14 & 9& 14& 14 & 8 & 7 \\
  -68.00&-242.94& 119.988548& -10.836110&3316& 0.30& 10.914&  0.855&  0.744&  0.174&  1.666 & 2.568 & 0.001& 0.001& 0.003& 0.004& 0.002& 0.002&G &   22& 14 & 9& 14& 14 & 8 & 7 \\
  -78.55& -25.95& 119.985916& -10.777542&3147&     & 10.931&  0.164&  0.206&  0.809&  0.433 & 2.799 & 0.002& 0.003& 0.003& 0.003& 0.004& 0.004&B &   22& 14 & 9& 14& 14 & 8 & 7 \\ 
  249.39&-190.99& 120.074715& -10.823801&    &     & 11.075&  1.012&  0.867& -0.034&  1.937 & 2.571 & 0.001& 0.002& 0.004& 0.007& 0.005& 0.004&G &   22& 14 & 9& 14& 14 & 8 & 7 \\
 -104.31& -51.21& 119.978897& -10.784216&3254& 0.20& 11.109&  0.886&  0.700&  0.080&  1.599 & 2.569 & 0.001& 0.002& 0.003& 0.004& 0.004& 0.003&G &   22& 14 & 9& 14& 14 & 8 & 7 \\
  242.51& 112.14& 120.073174& -10.742024&1375& 0.00& 11.252&  0.168&  0.148&  0.831&  0.293 & 2.779 & 0.001& 0.002& 0.003& 0.002& 0.003& 0.004&B &   23& 14 & 9& 14& 14 & 8 & 7 \\
 -148.05& 437.80& 119.967545& -10.652119&    &     & 11.373&  0.294&  0.133&  0.572&  0.403 & 2.675 & 0.001& 0.002& 0.003& 0.004& 0.003& 0.004&B &   23& 14 & 9& 14& 14 & 8 & 7 \\
 -359.63&  77.53& 119.909760& -10.748134&    &     & 11.596&  0.571&  0.387&  0.340&  1.015 & 2.574 & 0.000& 0.002& 0.004& 0.005& 0.004& 0.004&D &   23& 14 & 9& 14& 14 & 8 & 8 \\
   59.61& -29.52& 120.023399& -10.779244&2111& 0.00& 11.589&  0.366&  0.190&  0.380&  0.610 & 2.647 & 0.002& 0.003& 0.005& 0.004& 0.004& 0.005&B &   22& 14 & 9& 14& 14 & 8 & 7 \\
 -381.50&-154.26& 119.903580& -10.810519&    &     & 11.604&  0.563&  0.312&  0.432&  0.953 & 2.578 & 0.002& 0.002& 0.003& 0.002& 0.004& 0.003&G &   23& 13 & 9& 14& 14 & 6 & 6 \\
   61.83& -60.66& 120.023964& -10.787653&2122& 0.92& 11.726&  0.693&  0.409&  0.461&  1.211 & 2.571 & 0.001& 0.002& 0.004& 0.005& 0.004& 0.002&G &   23& 14 & 9& 14& 14 & 8 & 8 \\
 -462.34&-448.46& 119.881340& -10.889418&    &     & 11.745&  0.600&  0.333&  0.432&  1.003 & 2.574 & 0.002& 0.002& 0.002& 0.004& 0.004& 0.006&G &   23& 14 & 9& 14& 14 & 7 & 8 \\
  637.47& 359.90& 120.180595& -10.677331&    &     & 11.837&  0.753&  0.480&  0.480&  1.301 & 2.588 & 0.003& 0.004& 0.006& 0.006& 0.008& 0.006&G &   19& 11 & 9& 14& 10 & 5 & 5 \\
 -148.96&-453.43& 119.966362& -10.892436&    &     & 11.818&  0.361&  0.114&  0.364&  0.455 & 2.628 & 0.002& 0.003& 0.005& 0.006& 0.005& 0.005&B &   23& 14 & 9& 14& 14 & 8 & 8 \\
  -91.78& 317.66& 119.982689& -10.684816&4401&     & 11.840&  0.405&  0.197&  0.376&  0.670 & 2.620 & 0.001& 0.001& 0.003& 0.005& 0.003& 0.004&B &   23& 14 & 9& 14& 14 & 8 & 8 \\
 -307.53& 637.53& 119.924484& -10.597408&    &     & 11.870&  0.235&  0.085&  1.210&  9.999 & 9.509 & 0.004& 0.005& 0.009& 0.010& 9.999& 9.999&B &    3&  2 & 3&  5&  0 & 0 & 0 \\
  216.58&  79.61& 120.066101& -10.750658&1380& 0.00& 11.900&  0.358&  0.174&  0.450&  0.589 & 2.639 & 0.001& 0.002& 0.003& 0.004& 0.003& 0.003&B &   23& 14 & 9& 14& 14 & 8 & 8 \\
   74.87&   6.81& 120.027573& -10.769529&2101& 0.00& 11.957&  0.563&  0.236&  0.433&  0.863 & 2.573 & 0.002& 0.003& 0.004& 0.004& 0.004& 0.002&G &   23& 14 & 9& 14& 14 & 8 & 8 \\
  155.81& 172.77& 120.049713& -10.725212&1349& 0.00& 11.942&  0.307&  0.150&  0.419&  0.464 & 2.665 & 0.002& 0.003& 0.003& 0.003& 0.004& 0.005&B &   23& 14 & 9& 14& 14 & 8 & 8 \\ 
\enddata
\end{deluxetable}

With a final calibration in hand, it is possible to compare our data with published photometry in the cluster.
For $V$, we have matched our survey with the CCD photometry of \citet{MA97, KI01}. \citet{LE12} have done a large-area
survey of NGC 2506 down to $V$ $\sim$ 23.5 which encompasses our entire sample. While the data have been used to map
the structure and luminosity function of the cluster \citep{LE13}, it has never been published.

Using the WEBDA X,Y positions, we were able to cross-identify 576 stars in common with \citet{MA97} to $V$ = 18.0. 
The distribution in residuals, in the sense (Table 2 - MA97), showed no trends with $V$ or $b-y$, with virtually all 
the stars having residual differences less than 0.10 mag. If we exclude the 11 stars with absolute residuals greater than
0.10 mag, the average residual for the remaining 565 stars is 0.016 $\pm$ 0.028. Of the 11 deviants, three (5193, 5014, 2221)
show larger than average scatter in their indices for their $V$ magnitude and are likely suffering from PSF distortions
due to contamination by a nearby image. Star 5070 also has a very discrepant color compared to that expected from the
published $B-V$ and is a probable mismatch to the original survey. The remaining 7 discrepant stars are: 5233, 2121, 5312,
3147, 4238, 5176, and 2302. 

The comparison with the photometry of \citet{KI01} also revealed no statistically significant trends with magnitude or color,
but clearly suffered from larger photometric scatter than that of \citet{MA97}. The initial match of 536 stars brighter than $V$ = 
18 produced absolute residuals which were predominantly less than 0.20 mag. If all stars with differences greater than this
value are excluded, the remaining 515 stars show a mean offset of 0.017 $\pm$ 0.059, placing the \citet{KI01} $V$ on the
same system as \citet{MA97}. Because of the significantly larger scatter in the comparison with \citet{KI01}, we will not 
list the discrepant stars. 

Only one source of published $uvby$ photometry is available for comparison. \citet{AR07} make use of an unpublished CCD 
$uvby$ survey covering an area approximately $15\arcmin \times 15\arcmin$ to study and analyze variable stars within NGC 2506,
mostly $\delta$ Scuti and $\gamma$ Doradus variables, as well as a few eclipsing binaries.
While their survey is contained within the area of the current investigation and extends to $V \sim 19$, $uvby$ indices are
only published for the 28 variable candidates in the NGC 2506 field. Keeping in mind that all the stars under
discussion are variables, comparison of their Str\"omgren indices for 25 stars 
brighter than $V=17$ to ours leads to average differences, in the sense (Table 2 - AR07), and standard deviations for 
a single measurement for $V$, $b-y$, $m_1$ and $c_1$ are as follows: $+0.022 \pm 0.029, +0.006 \pm 0.010, 
-0.008 \pm 0.012$ and $+0.003 \pm 0.019$.  For two stars in Table 1, $m_1$ and $c_1$ values are absent, so information 
is based on only 23 stars for those indices. The agreement in zero-point and the small scatter among the residuals are
very encouraging and indicates that a more extensive comparison would show that the color indices of
both samples are on the same system to within 0.01 mag. 

\subsection{Variable Stars}
The hunt for variable stars in NGC 2506 using modern detectors begins with \citet{KA88}, who searched for long-period eclipsing
binaries within 6 overlapping fields covering a region of $6.2\arcmin \times 7.7\arcmin$,  down to $B \sim 18.7$. Not surprisingly, 
with only a few hours of observations, no eclipses were detected. As the field size and observing length have expanded with
each subsequent study, the number and types of variables have grown accordingly. \citet{KI01} surveyed a single field 
$5.8\arcmin \times 5.8\arcmin$, discovering four variables - one eclipsing binary and three $\delta$ Scuti variables. The 
latter variables were of particular interest because the blue straggler population and the top of the CMD turnoff of 
NGC 2506 sit within the predicted instability strip for this class of stars. Unexpectedly, comparable observations of 
NGC 2420 led to no variables, despite the similarity of the two clusters and a high estimated incidence of binaries, 
higher than in NGC 2506.

The most comprehensive survey to date is that of \citet{AR07}, covering a field $15\arcmin \times 15\arcmin$  with high precision and
$\sim$160 hours of observations each in $B$ and $I$. The study identified 28 variables in the cluster field, doubling 
the number of oscillating blue stragglers from 3 to 6 and the number of eclipsing variables from 1 to 2. More important, the
analysis identified 15 $\gamma$ Dor variables within the red hook of the turnoff of the CMD, in excellent agreement with the
predicted location from field stars of the same class \citep{HA02}. The value of the detected variables lies in the fundamental
stellar parameters they can supply through light curve analysis of the eclipsing variables and structural insights from
asteroseismology of the single-star variables.

While our CCD cluster frames have not been collected with a variable star search in mind, the relatively
high precision of the final photometry leaves open the possibility of identifying larger amplitude and/or longer period systems
through excessive dispersions among the indices or discrepant residuals in comparison with previous photometry. 
Building upon the first approach, all stars with $V < 16$ and a standard error of the mean in $b-y$ which implied larger than
typical scatter at a given $V$ (see Fig. 1) were identified. Stars with a reduced number of frames contributing to the index were
eliminated since this indicated that the star was near the edge of the CCD field or suffered severe contamination 
from a nearby object. Stars which showed significant scatter in only one filter but not both were removed, as were stars which showed
normal errors in filters other than $b$ and $y$, particularly H$\beta$. This left a sample of 25 potential variables. Each star
was then checked visually on a high quality $y$ frame, after which 14 stars were excluded due to asymmetric PSF's, implying contamination
by an optical companion. Of the 11 remaining stars, 6 lie within the area surveyed by \citet{AR07} and all 6 are known variables:
 V2, V3, V4, V9, V16, and V26. The list includes two oscillating blue stragglers, 3 eclipsing binaries, and 1 $\gamma$ Dor variable.
Given the success of the approach in identifying true variables within the well-studied cluster core, it is likely that the  
variables in the outer regions, members or not, are eclipsing binaries or stars within the main sequence instability strips. 
These stars are tagged within Table 2 by a V designation after the calibration class, i.e. GV or BV.

\section{The Color-Magnitude Diagram}
The CMD based upon ($V, b-y$) for all stars in Table 2 is shown in Fig. 2. Stars for which the internal errors in $b-y$
are below 0.015 mag are plotted as open circles. The five potential new variables are shown only in this figure as filled red squares 
because they lie outside the cluster core and will not be discussed in later analyses. The blue edge of the main sequence
CMD is well defined to $V$ $\sim$ 17.5, where increasing photometric scatter and the field star population with $b-y > 0.4$ begin
to dominate. The lack of scatter blueward of the main sequence is expected despite a distance of $\sim$ 3 kpc for the cluster
due to a galactic latitude near $-10$\arcdeg, coupled with a low line-of-sight reddening. The majority of contamination
will come from foreground stars with only slightly smaller reddening than the cluster. On the giant branch, with the 
exception of the obvious red clump, the separation of cluster members from the field is a challenge, particularly within 
the scatter of redder stars brighter than the clump.

\begin{figure}
\figurenum{2}
\plotone{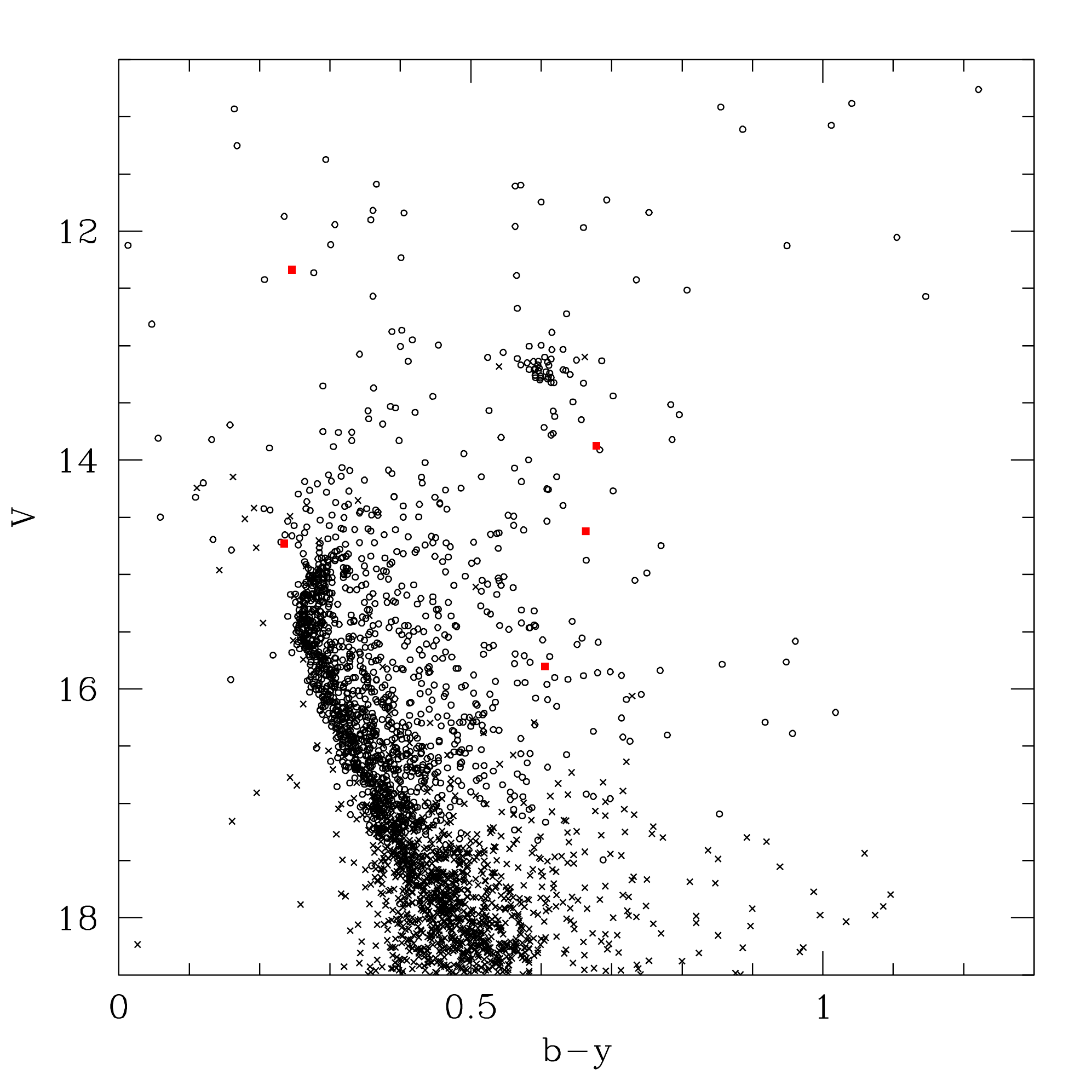}
\caption{CMD for all stars in Table 2. Filled red squares are new potential variables. Open circles are stars with errors in $b-y$ 
$\leqq$ 0.015 mag. Crosses are stars with errors greater than 0.015 mag.}
\end{figure}

As a first step toward isolating the cluster from the field, we restrict the sample to the cluster core. Using wide-field 
photometry \citep{LE12} on the $VI$ system, \citet{LE13} mapped the radial distribution of stars in NGC 2506, finding that 
the core radius varied as a function of the luminosity of the sample studied, ranging from 2.0\arcmin\ for the brighter 
stars to 6\arcmin\ for the lowest mass sample. For the entire dataset brighter than $V = 23$, they found a core radius 
of 4\arcmin, which we will adopt for our core selection, shown in Fig. 3, where symbols have the same meaning as in Fig. 2. 
The improvement is striking. While the unevolved main sequence is again obvious, the rich sample of stars above the turnoff 
between $V = 13.7$ and 14.7 is almost certainly populated by binaries and a small sample of single stars evolving from the 
red hook to the subgiant branch. The scatter of stars redward of the main sequence is severely reduced and most of the 
scatter lies within 0.75 mag of the main  sequence, implying a pattern defined by binaries.
The subgiant branch is sparsely populated, but the path of the first-ascent red giants starting at the base near 
$V \sim  15$ is discernible. Equally important, the majority of the red giants above the clump are gone, an 
indication that most are likely non-members.

\begin{figure}
\figurenum{3}
\plotone{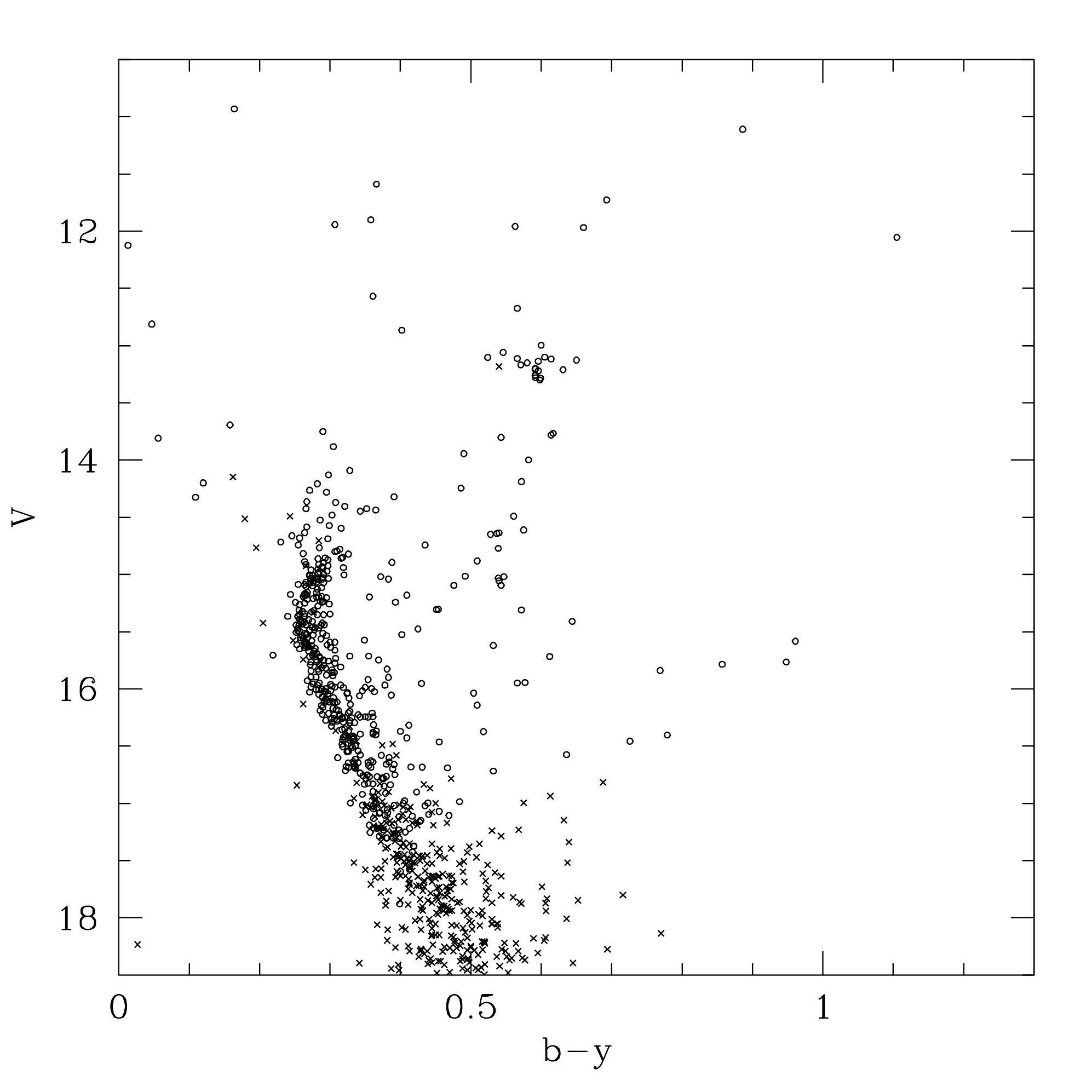}
\caption{CMD of stars within $4\arcmin$ of the cluster core. Symbols have the same meaning as in Fig. 2.}
\end{figure}

The dataset of Fig. 3 would be a solid starting point for analysis of the cluster parameters if no additional information were
available. However, NGC 2506 has been the subject of a proper-motion analysis \citep{CA81} and 34 red giants in the cluster
field have been surveyed for radial-velocity variations \citep{ME07, ME08}. We can therefore expand the probable cluster sample
beyond the core, eliminate likely non-members throughout the CCD field, and cross-check the proper-motion membership for the 
giants against the radial-velocity predictions. The result of this more complicated approach is shown in Fig. 4 where only stars
with errors in $b-y$ $\leq$ 0.015 have been retained. Note the reduction in the $b-y$ and $V$ range for the figure. Black 
symbols denote stars within the cluster core with membership probabilities above 50\%, while red symbols show stars that meet the same criterion
but which fall outside the core.  Green symbols are stars within the core for which no proper-motion data are available, and 
crosses, red or black, are red giants with radial velocities consistent with the very high cluster value but non-zero 
proper-motion probabilities below 50\%. As expected, the green points are responsible for the majority of the deviant 
points fainter than the subgiant branch, but the cluster members fall consistently within the band defined by the main 
sequence and the brighter limit of the binary sequence.

\begin{figure}
\figurenum{4}
\plotone{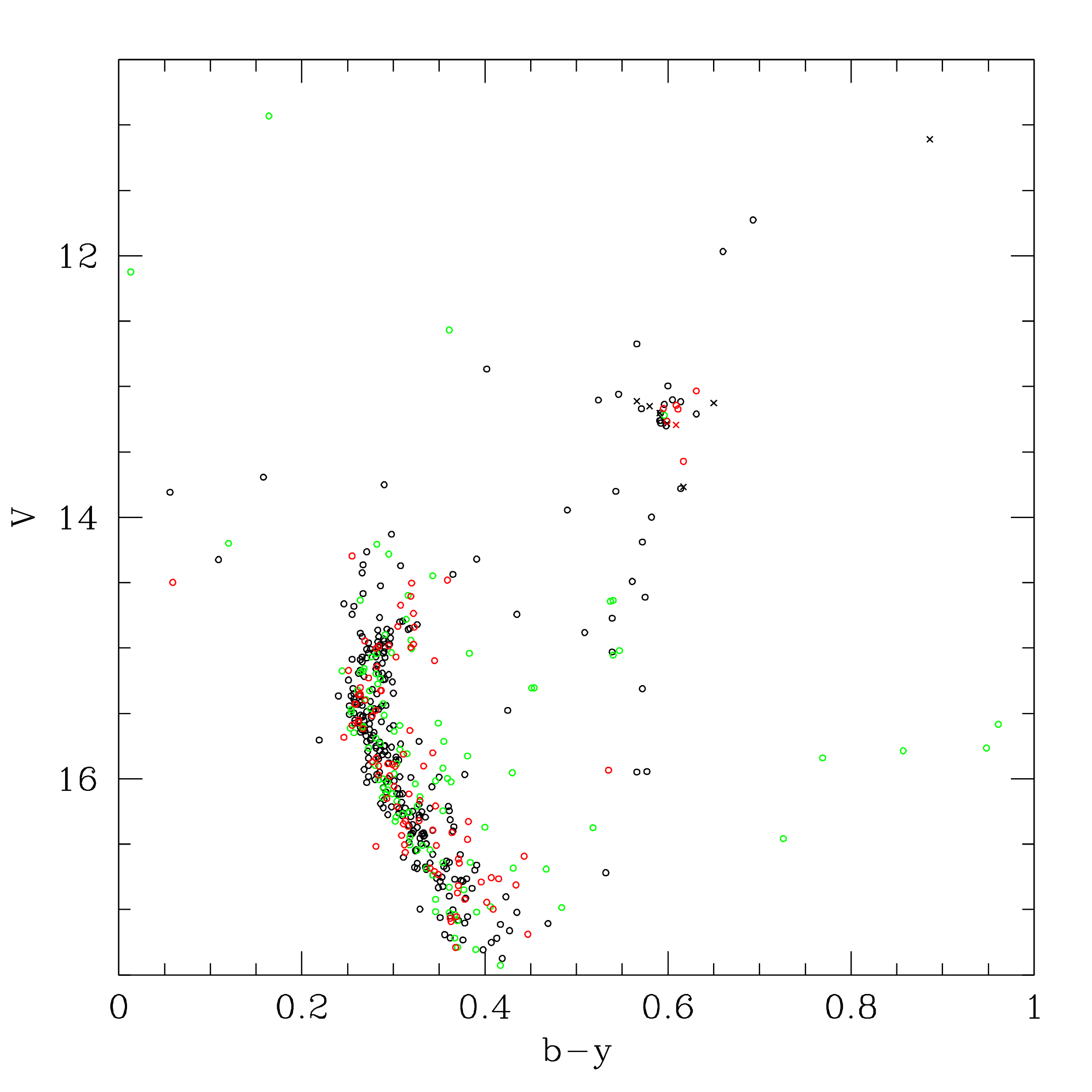}
\caption{CMD for stars with errors in $b-y$ $\leqq$ 0.015 mag proper-motion membership above 50\%  within the core (black circles), 
outside the core (red circles), inside the core but no proper-motion information (green circles), and radial-velocity members with
proper-motion membership below 50\% (crosses).}
\end{figure}

\section{Reddening and Metallicity}
With a base of probable cluster members with high-precision $by$ photometry in hand, we shift our focus to the stars at the cluster
turnoff where the color range is optimally suited for analysis on the extended Str\"omgren system. To reduce any
potential distortions caused by evolution well off of the main sequence and contamination by highly probable binaries which
populate the vertical band of stars above the turnoff between $V$ = 13.75 and 14.75, we eliminate any stars brighter than
$V$ = 14.8. We then eliminate all stars which have photometric errors in $m_1$, $c_1$, $hk$, and H$\beta$ greater than
0.020, 0.030, 0.020, and 0.015 mag, respectively. This imposes an effective cutoff on the sample near $V$ = 16.70. Finally,
we narrow the color range to stars with $b-y$ between 0.20 and 0.60, leaving a sample of 332 stars, plotted in Fig. 5.
While some stars scatter well to the red of the main sequence because they are either field stars or giants near the base
of the giant branch, the dominant scatter appears to be caused by the cluster binary sequence extending parallel to but
brighter than the normal turnoff. Given the relatively high precision of the $b-y$ indices, we can eliminate a large
fraction of these stars by tagging any stars which lie off the mean relation by more than $\sim$0.035 mag. 
Stars plotted as red crosses in Fig. 5 have been tagged as meeting this criterion. A pair of stars which lie blueward of the mean
relation are plotted as blue squares. Note that the number of red points effectively disappears near the turnoff above $V$ = 15.6;
the binary sequence should merge with the single-star sequence near this region of the CMD, making identification of the
binaries impossible using this simple technique.

\begin{figure}
\figurenum{5}
\plotone{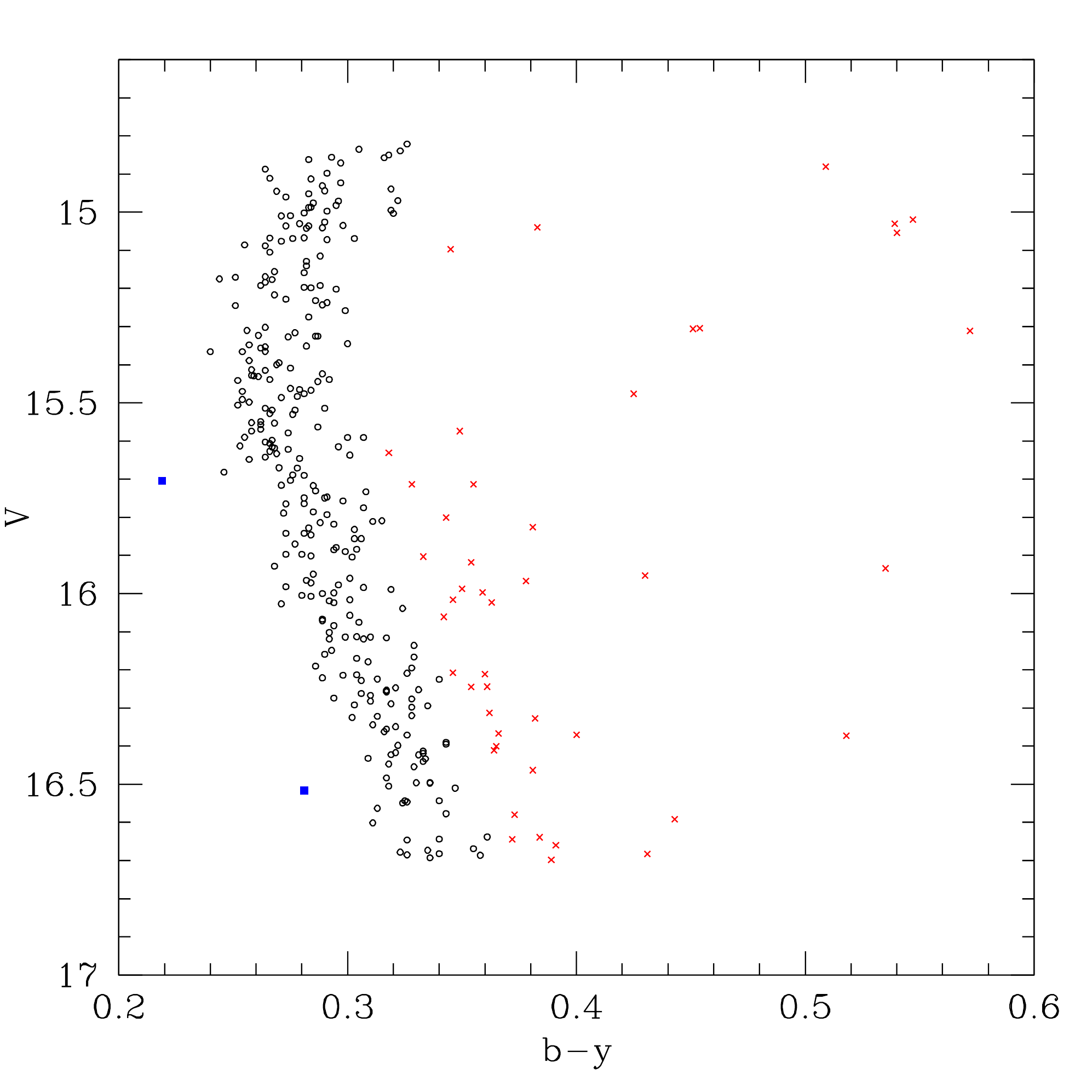}
\caption{CMD of stars from Fig. 4 with precision cuts to eliminate stars with larger errors in any and all indices. Stars tagged as
deviants from the mean cluster relation are plotted as blue squares and red crosses to establish which side of the turnoff region they
occupy.}
\end{figure}

We now turn to the $V, hk$ CMD as a test of our sample selection and as a means to eliminate additional non-members. 
The use of $hk$ is valuable for three reasons. First, for a fixed
metallicity, $hk$ is much more sensitive to temperature changes than $b-y$ for mid-F stars and cooler. Stars which are truly 
cooler at a given $V$ compared to the mean cluster relation, either because they are binaries shifted vertically relative to
single stars in the CMD or field stars, will exhibit a measurable shift in $hk$. Second, $hk$ is one-sixth as sensitive to
reddening as $b-y$ and shifts the star to bluer/smaller $hk$ values. Stars which are unusually reddened in the $V, b-y$ CMD
will remain virtually unchanged  or appear bluer in the $V, hk$ diagram. Third, of special importance to NGC 2506 where the 
turnoff stars are late A/early F spectral types with very similar $hk$ indices over a range in $b-y$, $hk$ is extremely 
sensitive to small changes in metallicity. NGC 2506 is predicted to be metal-deficient by 0.2 to 0.3 dex relative to the sun. 
If the foreground  field stars along the line of sight are typical of the solar neighborhood, they should easily separate 
from the cluster sample by an amount which is three to five times larger than the photometric errors. 

\begin{figure}
\figurenum{6}
\plotone{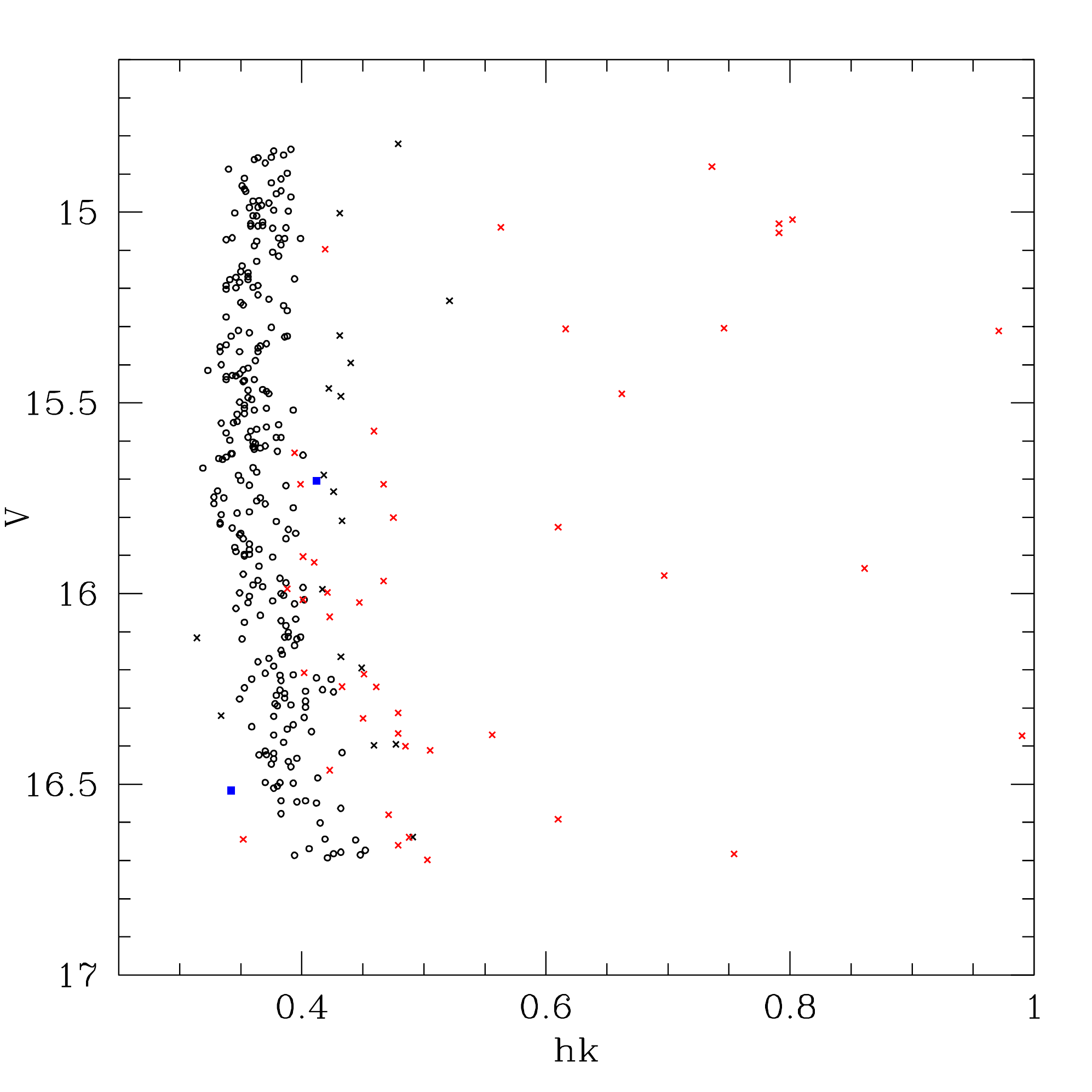}
\caption{CMD in $V, hk$ for the stars of Fig. 5. Symbols have the same meaning as in Fig. 5, with the addition of black
crosses to identify stars which deviate in $hk$ at a given $V$ not tagged as deviants in Fig. 5.}
\end{figure}

Fig. 6 shows the CMD with the same symbols as Fig. 5, plus one additional class. Black crosses are stars which deviate 
significantly from the cluster sequence in $hk$ at a given $V$ magnitude which were not tagged in Fig. 5. The almost
vertical band defined by the hot stars at the turnoff of NGC 2506 is readily apparent, with a scatter at a fixed $V$
consistent with the restricted photometric error limits attached to the sample selection.
As hoped, the general pattern is that the stars which lie redward of the cluster sequence in Fig. 5 (red crosses) reproduce a 
similar trend in Fig. 6. There are, however, 18 stars which were not identified as being deviant in Fig. 5 which appear 
to be so in Fig. 6; the majority of these are possible binaries or, more likely, field stars which have significantly 
higher [Fe/H] than the cluster but happen to superpose on the cluster $V, b-y$ CMD due to a difference in absolute magnitude.
It should be noted that for a metal-deficient cluster, a similar selection criterion could be built upon the $m_1$ index, with
metal-rich field stars having a larger $m_1$ value at a given $V$. For comparable photometric precision, however, the 
expected change in $m_1$ would be 3-4 times smaller, making it difficult to distinguish between the impact of metallicity and 
photometric errors.

With all stars plotted as crosses in Fig. 6 eliminated, the photometric reddening and metallicity can be derived. For details
on the approach, the reader is referred to a subset of the several past investigations where it has been applied \citep{AT14, TW15, 
SA16}. The simultaneous solution for [Fe/H] and $E(b-y)$ based upon $m_1$ and H$\beta$ for 257 stars is $-0.296 \pm 0.011$ (sem) 
and $0.042 \pm 0.001$ (sem) ($E(B-V) = 0.058 \pm 0.001$, respectively, independent of the standard relation adopted for 
the Str\"omgren indices \citep{OL88, NI88}. With the reddening set, the metallicity based upon $hk$, H$\beta$ is 
[Fe/H] $= -0.317 \pm 0.004$ (sem). For a change in $E(b-y)$ of $\pm 0.010$ ($E(B-V) = \pm 0.014$), 
[Fe/H]$_{m1}$ changes by $\pm 0.034$ while [Fe/H]$_{hk}$ changes by only $\pm 0.006$ dex. For an uncertainty 
of $\pm 0.009$ in the zero-points of the $m_1$ and $hk$ indices, the changes in [Fe/H] are $\pm 0.122$ and $\pm 0.032$, 
respectively. The weighted mean of the two abundances is clearly dominated by $hk$, leading to [Fe/H] $= -0.316 \pm 0.033$, where 
the error estimate includes both internal and potential systematic errors.

\subsection{Red Giants and Reddening}
The red giants are often ignored in discussions of the reddening on the traditional Str\"omgren system because H$\beta$ becomes
insensitive to changes in temperature among cooler stars, approaching a limiting index near 2.56 for increasing $b-y$ (see Fig. 2 
of \citet{TW15}). With the addition of $hk$, the giants take on added significance because of the very weak reddening
dependence of the $hk$ index coupled to a temperature dependence which is 2.5 times more sensitive than $b-y$. To illustrate
the point, in Fig. 7 we plot all the red giants of Fig. 4 with the same color (black) irrespective of location in
the cluster field; black crosses are still giants identified as probable members based upon radial velocity, independent of the
proper-motion membership probability. The blue line is the relation drawn defined only by the cluster members. The residuals about the
mean relation have a dispersion of $\pm$0.010 mag. From the quoted photometric errors, the dispersion is predicted to be $\pm$0.004
$\pm$ 0.002 mag. Since all the stars should have the same metallicity, and reddening effects on $hk$ are six times smaller and in
the opposite direction to the effects on $b-y$, the dispersion in $b-y$ at a given $hk$ should provide an indication of the
size of the reddening spread across the face of the cluster. The upper limit on the spread in $E(b-y)$ becomes $\pm 0.009$ mag.
Considering that the quoted errors on the photometry are likely lower limits, that binaries and other giants with potential 
spectroscopic anomalies have not been eliminated, that giants at a given $hk$ may be first-ascent, red-clump, or asymptotic
giant branch stars where surface gravity changes may affect the color indices, we place a conservative upper limit to 
the reddening variation across the face of the cluster at $\pm$0.007 for $E(b-y)$ or $\pm$0.01 $E(B-V)$, i.e. effectively 0.0.

\begin{figure}
\figurenum{7}
\plotone{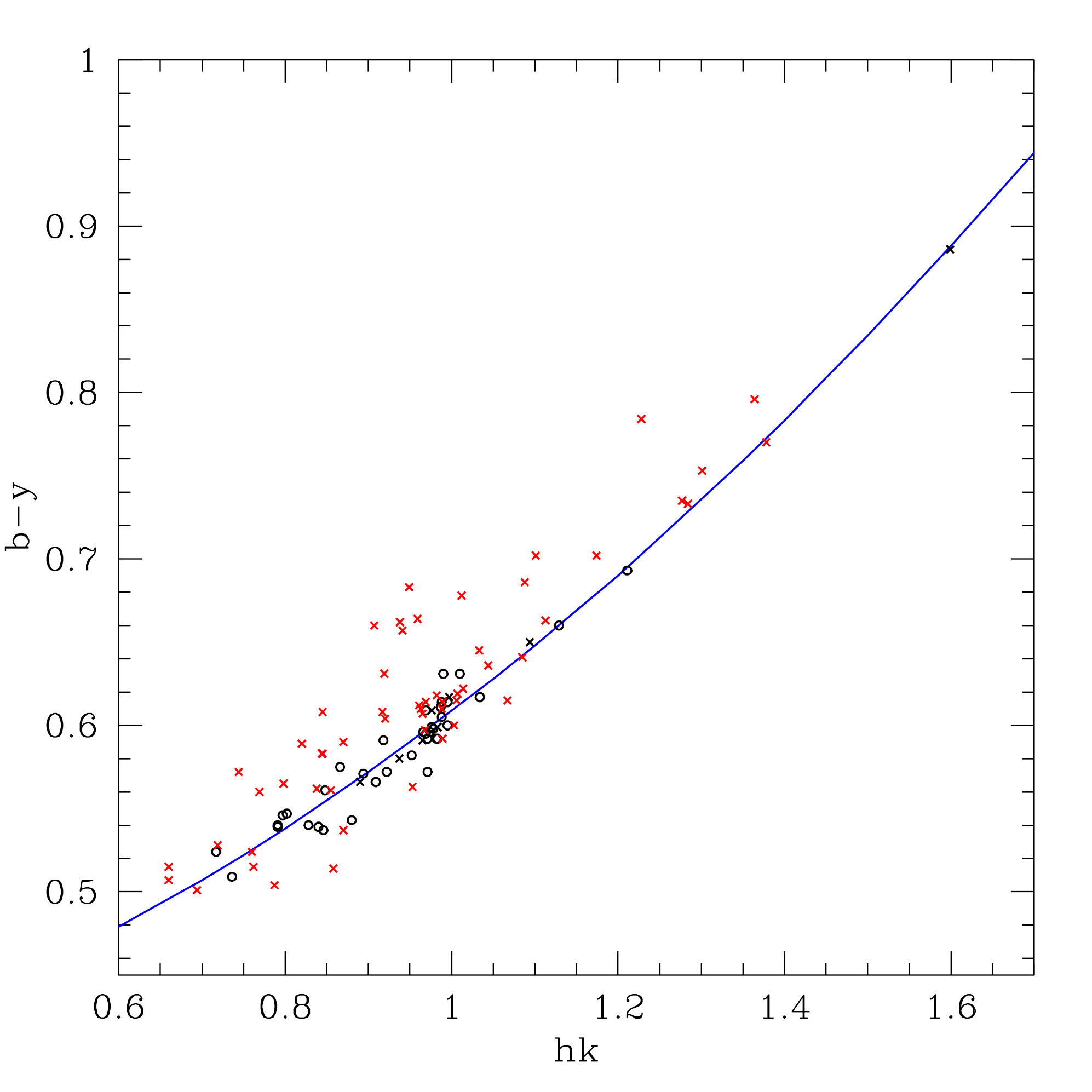}
\caption{$b-y$, $hk$ relation (blue line) based upon probable cluster members (black circles and crosses). Red crosses
are stars outside the cluster core without proper-motion or radial-velocity measurements.}
\end{figure}

With the relation for cluster members well-defined, we can now add to Fig. 7 all stars outside the cluster core for which proper-motion and
radial-velocity information is absent (red crosses). The large scatter of points away from and above the relation for the cluster
confirms that the majority of these are field giants, likely background giants with higher reddening and likely lower metallicity
than the cluster. There are, however, approximately two dozen stars which have positions consistent with cluster parameters, dominated 
by a rich concentration of a dozen stars at the colors of the clump ($hk, b-y)$ $\sim$ (1.0, 0.6). Further observation and
analysis of these candidates could add insight to the already rich red giant population of NGC 2506.

\subsection{Age and Distance}
Our entire analysis is based upon Str\"omgren photometry, so for consistency with our recent analysis of NGC 752 \citep{TW15},
we turn to the Victoria-Regina isochrones \citep{VA06} for simultaneous age and distance estimation. Fig. 8 shows the data
of Fig. 4 with the isochrones for ages 1.7, 1.8, and 1.9 Gyr and [Fe/H] $= -0.29$, the closest abundance to the derived metallicity of
the cluster, superposed. The adopted best-fit, apparent modulus, assuming $E(b-y)$ = 0.042 is $(m-M)$ = 12.75. The matches to the
turnoff, subgiant, and giant branches are very good to excellent for an age between 1.8 and 1.9 Gyr. For the giant branch,
it is intriguing to note that while the sequence below the level of the clump fits the observations well, the few stars brighter
than the clump all lie at significantly brighter positions than implied by the isochrones, leading to the conclusion that these
stars are binaries and/or asymptotic giant branch members of the cluster.

\begin{figure}
\figurenum{8}
\plotone{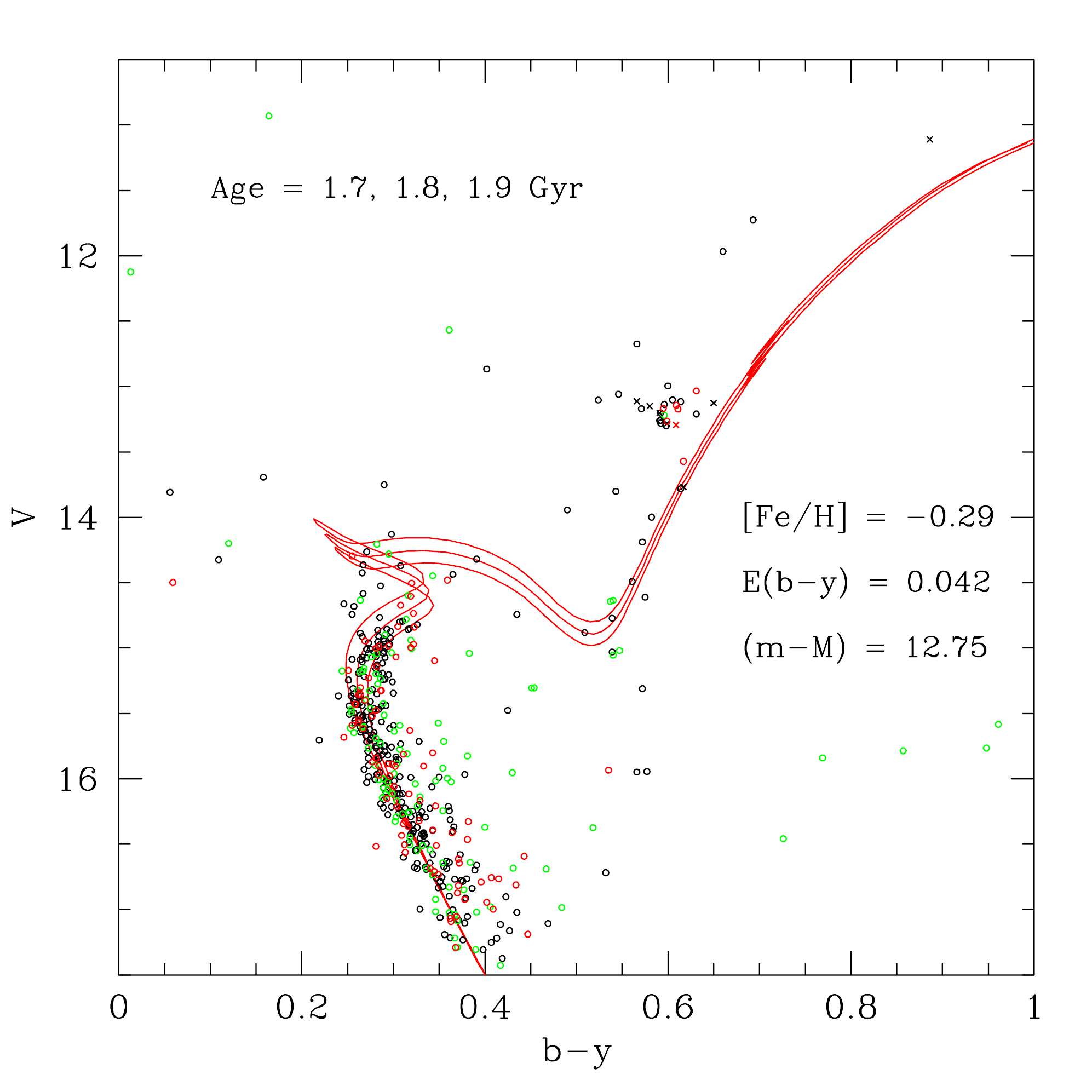}
\caption{CMD of Fig. 4 with Victoria-Regina isochrones of ages 1.7, 1.8, and 1.9 superposed.The adopted reddening and the
apparent distance modulus are $E(b-y$ = 0.042 and $(m-M)$ = 12.75, respectively.}
\end{figure}

The impact of the difference
between the true [Fe/H] and the isochrone value is small enough to ignore for the age estimate, though a more metal-poor
isochrone would produce a slightly older age. With respect to the distance modulus, the more metal-rich main sequence overestimates the modulus by less than
0.03 mag, well within the uncertainties posed by variations in the reddening and the transformation of the isochrones from
the theoretical plane to the observational.

\subsection{Comparison With Previous Results}
Since the direct measurement of $E(B-V)$ for NGC 2506 using DDO photometry of 7 red giants in the field of the cluster 
\citep{MC81} produced 0.052 $\pm$ 0.016, it has been clear that the cluster reddening is small. The reddening maps 
of \citet{SC11}, revising the earlier work of \citet{SC98}, produce $E(B-V)$ = 0.074 in the direction of the cluster, confirming 
a modest reddening even at large distance, primarily due to the large galactic latitude. Later investigations have 
adopted $E(B-V) = 0.05$ to 0.08 \citep{TW97, CA94, FR02}, derived values between 0.0 and 0.07 using isochrone fits 
which depend on the adopted [Fe/H] \citep{MA97, LE12}, or made direct determinations from multicolor 
photometry ($E(B-V) = 0.04 \pm 0.03$, \citet{KI01}). \citet{CA04} used spectroscopic temperatures 
coupled with photometry to obtain $E(B-V) = 0.073 \pm 0.009$ from 4 giants, a value adopted by \citet{RE12} and \citet{AR07}. 
In summary, accounting for NGC 2506's metal deficiency, the previously published values
obtained by spectroscopic, photometric, and/or CMD analysis are consistent with $E(B-V)$ = 0.055 $\pm$ 0.020, in agreement  
with the value derived here. Equally important, changing the reddening by $E(b-y)$ = $\pm$0.015 ($E(B-V)$ = $\pm$0.02) would
alter the derived cluster age by less than $\mp$0.1 Gyr.

Metallicity comparisons can be challenging given the often different assumptions for reddening and for the zero-point of the
metallicity scale upon which the measurement is based. The first direct metallicity estimate for NGC 2506 was based on a 
combination of DDO red giant photometry and $UBV$ photometry of both red giants and turnoff stars, leading to 
[Fe/H] $= -0.55$ {\it relative to the Hyades}, which defined the standard relations
adopted in the analysis \citep{MC81}. If [Fe/H] for the Hyades is assumed to be +0.12, this implied a cluster mean of $-0.43$. We note that values as high as +0.18 have been derived in recent studies (e.g., \citet{Dutra} and references therein).
For reference, a virtually identical approach using the same calibration relations on NGC 752 \citep{TW83} led to 
[Fe/H] {\it relative to the Hyades} of $-0.35$, making it 0.20 dex more metal-rich than NGC 2506. The DDO system was later 
recalibrated \citep{TW96} but, not surprisingly, while the absolute scale shifted slightly, the differential offset 
remained basically the same, with [Fe/H] $= -0.37$ and $-0.16$ for NGC 2506 and NGC 752, respectively \citep{TW97}. The first 
spectroscopic work on  NGC 2506 was the moderate-resolution spectroscopy of
\citet{FJ93}, who derived [Fe/H] $= -0.52$ on a scale where NGC 752 had [Fe/H] $= -0.16$. This data set
was revisited and recalibrated by \citet{TW97}, who derived [Fe/H] $= -0.38$ and $-0.09$ for NGC 2506 and NGC 752.
The moderate-resolution spectroscopic data were expanded and recalibrated by \citet{FR02}, producing [Fe/H] $= -0.44$ and $-0.18$ for
NGC 2506 and NGC 752. The often overlooked aspect of this final revision is that the definition of the metallicity 
scale included the clusters M67 and NGC 2420 as standards, assumed to have input [Fe/H] $= -0.10$ and $-0.42$, respectively, and 
generating derived values for these two clusters of $-0.15$ and $-0.38$. The clear consensus from the photometric and moderate-dispersion
spectroscopy is that NGC 2506 lies 0.20 to 0.30 dex below NGC 752 in [Fe/H]. 

As mentioned in the Introduction, high-dispersion spectroscopic abundances within NGC 2506 are few and far between. \citet{CA04}
analyzed 4 giants with an unweighted average of [Fe/H] $= -0.24 \pm 0.09$ (sd). Restricting the sample to only two clump giants, 
this became [Fe/H]$= -0.20 \pm 0.02$ (sd). The same four stars were reanalyzed by \citet{MI11}, 
leading to [Fe/H] $= -0.24 \pm 0.05$ (sd). Finally, the spectroscopic work of \citet{RE12} generated [Fe/H] $= -0.19 \pm 0.06$
for 3 giants in NGC 2506 on a scale where 4 giants in NGC 752 led to [Fe/H]$ = -0.02 \pm 0.05$.

For age and distance, we restrict our comparisons to values derived via direct comparison of cluster photometry to isochrones of
appropriate vintage and metallicity. \citet{NE16}, e.g., quote a mean age of 1.625 $\pm$ 0.429 from 20 age determinations
published in the literature. Unfortunately, a check of the potential sources for these age estimates shows that most are
outdated, of poor quality, and/or repetitions of values from other catalogs. An often-repeated source for the 
age of NGC 2506 is the listing in WEBDA at 1.1 Gyr (log t = 9.045), despite the fact that no isochrone fit to the
cluster has at any time in the last 35 years generated an age this small. This anomalously low value has been propagated through
other catalogs such as \citet{DI02, TA02, KH05}, just to name a few. Even using basic CMD morphological parameters 
\citep{TW89, JD83, JP94, CA94, SA04} to determine age, NGC 2506 always occupies a position intermediate in age between NGC 752 and M67 and
above 1.5 Gyr. 

Following the original discussion by \citet{MC81}, the first attempt to determine the age of NGC 2506 using isochrones incorporating
convective overshoot mixing is that of \citet{CA94}. With $E(B-V)$ and [Fe/H] set at 0.08 and $-0.52$, they derived an age
of 1.9 Gyr and an apparent modulus of $(m-M) = 12.75$. If we lower the reddening to 0.058 and raise the metallicity to $-0.32$,
the effects on the distance modulus mostly cancel, leading to an increase in $(m-M)$ to just above 12.8. Likewise, lowering the
reddening increases the age while raising the metallicity causes it to drop; without more details on the specific isochrones, the
net change is impossible to estimate precisely. \citet{MA97} explored a range of models and cluster parameters, where the exact
values of optimum distance, reddening, age and metallicity varied with the isochrones under discussion. However, the optimal
set of ages and distances ranged from 1.5 to 2.2 Gyr and $(m-M)_{0} = 12.5$ to 12.7 for all plausible choices of reddening 
and metallicity. 

\citet{TW99} used $E(B-V)$ = 0.04 and isochrones with convective overshoot mixing at [Fe/H] $= -0.39$ to derive $(m-M) = 12.7$ 
and an age of 1.9 Gyr. Measuring $E(B-V) = 0.03$ and using isochrones with [Fe/H] $= -0.4$, \citet{KI01} found an age of 1.8 Gyr and
and apparent modulus of $(m-M) = 12.6$. In both cases, raising the reddening makes the cluster slightly younger and 
increases the apparent modulus by 0.1 and 0.15 mag, respectively. The higher metallicity would also boost the distance 
modulus and slightly decrease the age.  

Finally, \citet{LE12} have used unpublished $VI$ photometry to obtain an optimal isochrone match for [Fe/H] $= -0.24$, $E(B-V) = 0.035$,
and $(m-M) = 12.58$ and an age of 2.31 Gyr. Raising the reddening and lowering the metallicity will lead to opposite effects on the
age and distance modulus. No plausible shift in metallicity and/or reddening can bring this age estimate into agreement with the
results of Fig.7.

\section{Summary}
NGC 2506 is an older open cluster which has long been classed as metal-deficient. The exact age and metallicity, critical for
coherent interpretation of the evolutionary trends among the stars as revealed by their surface parameters and elemental abundances,
have generally converged over the last 20 years toward an age between 1.7 and 2.0 Gyr and a metallicity between 0.2 and 0.3 dex
less than the better-studied cluster NGC 752. Precision $uvbyCa$H$\beta$ CCD photometry covering a field $20\arcmin \times 20\arcmin$,
when coupled with membership information from proper motions and radial velocities, allows reliable isolation of the cluster 
from the field. Restricting the members to 257 turnoff stars within the ideal calibration range for the extended Str\"omgren system,
the reddening is calculated as $E(b-y) = 0.042$ ($E(B-V) = 0.058$), in excellent agreement with the range of published values
from a variety of techniques. Moreover, contrary to past suggestions tied to inadequate photometry and membership information,
there appears to be no evidence for significant variation in $E(B-V)$ across the face of the cluster at a level greater
than $\pm 0.01$ mag. The estimated metallicities from $m_1$ and $hk$ are $-0.29$ and $-0.32$, on a scale where
NGC 752 has [Fe/H] $= -0.07$ and $-0.02$, respectively, again demonstrating that the cluster is typically 0.25 dex more metal-poor
than NGC 752. Giving more weight to the $hk$-based metallicity due to its lower sensitivity to reddening changes and to zero-point
uncertainties than $m_1$, isochrones with $E(b-y) = 0.042$ and [Fe/H] $= -0.29$ define a cluster age of 1.85 $\pm 0.05$ Gyr
for an apparent modulus of $(m-M) = 12.75 \pm 0.1$. With this age and metallicity, NGC 2506 occupies a key parameter space
for mapping the impact of age and metallicity on the evolution of Li among stars of intermediate mass, comparable in age to
NGC 3680 (1.75 Gyr) and IC 4651 (1.5 Gyr), but significantly more metal-poor than NGC 3680 ($-0.08$) and IC 4651 ($+0.12$)  
\citep{AT09}. The addition of two dozen potential red giant members of the cluster outside the core will permit
a more detailed delineation of the effects of post-main-sequence evolution on the atmospheric Li, a point we will return 
to in a future discussion of our high-resolution Li spectroscopy.

\acknowledgments

The authors gratefully acknowledge extensive use of the WEBDA\footnote{http:// webda.physics.muni.cz} database, 
maintained at the University of Brno by E. Paunzen, C. Stutz and J. Janik, and 
TOPCAT\footnote{http://www.star.bris.ac.uk/$\sim$mbt/topcat/}. Filters used for these observations 
were obtained through NSF grant AST-0321247 to the University of Kansas. Continuing support has 
been provided to BJAT and BAT through NSF grant AST-1211621 and to CPD through AST-1211699.

\facility{CTIO:1.0m}
\software{IRAF}

\end{document}